\documentclass[pdftex,twocolumn,epjc3]{svjour3}          

\RequirePackage[T1]{fontenc}

\smartqed  
\usepackage{amsmath, latexsym}
\RequirePackage{graphicx}
\RequirePackage{mathptmx}      
\RequirePackage{flushend}
\RequirePackage[numbers,sort&compress]{natbib}
\RequirePackage[colorlinks,citecolor=blue,urlcolor=blue,linkcolor=blue]{hyperref}

\journalname{Eur. Phys. J. C}

\begin{document}

\title{Self Similar Collapse and the Raychaudhuri equation}


\author{Shibendu Gupta Choudhury\thanksref{e1,addr1}
        \and
        Soumya Chakrabarti\thanksref{e2,addr2} 
        \and Ananda Dasgupta\thanksref{e3,addr1}
        \and Narayan Banerjee\thanksref{e4,addr1}
}

\thankstext{e1}{e-mail: sgc14ip003@iiserkol.ac.in}
\thankstext{e2}{e-mail: soumya.chakrabarti@saha.ac.in}
\thankstext{e3}{e-mail: adg@iiserkol.ac.in}
\thankstext{e4}{e-mail: narayan@iiserkol.ac.in}
\institute{Department Of Physical Sciences, IISER Kolkata, Mohanpur, Nadia 741246, India\label{addr1}
          \and
          Theory Division, Saha Institute of Nuclear Physics, Kolkata 700064, India\label{addr2}}

\date{Received: date / Accepted: date}

\maketitle

\begin{abstract}
The role of the Raychaudhuri equation in studying gravitational collapse is discussed. A self-similar distribution of a scalar field along with an imperfect fluid in a conformally flat spacetime is considered for the purpose. The general focusing condition is found out and verified against the available exact solutions. The connection between the Raychaudhuri equation and the critical phenomena is also explored. 
\end{abstract}
\section{Introduction}
It is widely accepted that at the end of its life-cycle, a massive astronomical body undergoes a phase of gravitational collapse. 
 Followed by the pioneering work on an idealized model star collapsing to zero
volume by Datt \cite{datt} and Oppenheimer and Snyder \cite{os}, many attempts have been made to find a more complete and increasingly 
generalized description of gravitational collapse. For a detailed review of the relevance 
of gravitational collapse in physics and the open problems, we refer to the references \cite{joshi1, joshi2}. \\

 It is not a trivial task to simplify the Einstein field equations enough to guarantee the extraction of an exact solution that can describe the 
collapsing evolution. This often compels one to resort to the study of simplified systems which can describe the essential physics. 
 Christodoulou studied the global initial value problem for Einstein's equations in the spherically symmetric case with a massless 
scalar field and showed that the scalar field, depending on the initial 
collapsing profile, can either converge towards a zero proper volume or bounce, dispersing towards infinity \cite{christo, christo2, christo3}. Choptuik studied a similar problem numerically and pointed out the now well known critical 
phenomena in massless scalar field collapse \cite{chop}. Similar critical phenomena were studied by 
Brady, Chambers and Goncalves \cite{bcg} and Gundlach \cite{gundprl}. Throughout the last decade, many attempts have been made to 
generalize the ideas of critical phenomena for less simplified scenarios. For example, critical phenomena in a collapsing system with 
angular momentum was studied by Olabarrieta, Ventrella, Choptuik and Unruh \cite{ola}. The critical behavior of a spherically symmetric 
collision-less matter at the threshold of black hole formation was studied by Olabarrieta and Choptuik \cite{olachop}. Analytical 
investigations of the critical collapse problem, and a search for a theoretical explanation for the behavior discovered by Choptuik were 
carried out by Brady \cite{brady}, under the assumption that the collapse is self-similar. For a detailed review on the critical phenomena in Gravitational Collapse we refer to the review by 
Gundlach and Martin-Garcia \cite{Gundlach1}. \\

 The current work stems from  a different motivation -  we intend to investigate the role of the Raychaudhuri equation \cite{rc, ehlers}, 
in the context of the study of gravitational collapse and critical phenomena. 
 This equation has been used to study relativistic charged collapse by Kouretsis and Tsagas \cite{kt}. 
 In the present work we study the evolution of a conformally flat geometry, minimally  coupled with a scalar field, 
along with the presence of a fluid.
This specific case of spacetime metric is well-studied for radiating and/or
shear-free stars \cite{som, maiti, modak, bhui, patel, schafer, ivanov, herrera} and is receiving increasing interest in the 
context of gravitational collapse quite recently \cite{hamid, scnbepjc, scnbprd}.  
 We make the additional assumption that the evolution
is self-similar in nature.
Gravitational collapse of a self-similar fluid with heat
flow in a conformally flat spacetime was studied by Chan, Silva and Rocha \cite{chan0}.
Collapsing models of fluids with the assumption of self-similarity in more general spacetimes were studied
in \cite{brandt,chan1,chan2,chan3} and references therein. In this work,  we first discuss the usefulness of the Raychaudhuri equation
in studying the dynamics of the system. Generic conditions regarding the evolution of the spacetime can be provided using the Raychaudhuri
equation. These conditions are useful in the absence of an exact solution. We then work out 
exact solutions of field equations for some special cases making use of some simplifying assumptions. We  show  the consistency of the conclusions  regarding the dynamics from  exact solutions, if available,
 with those arrived at from the Raychaudhuri equation.
The role of the Raychaudhuri equation in the context of the critical phenomena
is also studied. \\

The paper is organized as follows.  In section \ref{sec2}
we introduce our system and discuss the assumptions involved. The field equations and conservation equations for the system are also presented in this section.
The Raychaudhuri equation and focusing condition are briefly discussed in the section \ref{sec3}. In this section, 
we focus on the role of the Raychaudhuri equation in extracting information about
 the evolution of  spacetime. This section also includes a discussion on the consistency of the results with exact solutions of the field
 equations, where available, and a connection between focusing condition and the critical phenomena. The final section {\ref{sec4}} includes some concluding remarks.

\section{The System}\label{sec2}
The spacetime which we consider is a conformally flat spherically symmetric spacetime for which the metric can be written as,  
\begin{equation}
\label{metric}
ds^2=\frac{1}{{A^2 (r,t)}}\Bigg[-dt^2+dr^2+r^2d\Omega^2\Bigg],
\end{equation}
where $\frac{1}{A^2 (r,t)}$ is the conformal factor which governs the evolution of the 2-sphere.        

The contribution to the energy-momentum tensor comes from a scalar field and a fluid,
\begin{equation}\label{EMT}
T_{\mu\nu}=T^\phi_{\mu\nu}+T^{\mathrm{fluid}}_{\mu\nu},
\end{equation}
where \begin{equation}
     T^\phi_{\mu\nu}=\partial_\mu\phi\partial_\nu\phi-g_{\mu\nu}\Bigg[\frac{1}{2}g^{\alpha\beta}\partial_\alpha\phi\partial_\beta\phi+
V(\phi)\Bigg],
    \end{equation} 
and \begin{equation}
\begin{split}
       T^{\mathrm{fluid}}_{\mu\nu}=&\left(\rho+p_t\right)u_\mu u_\nu+p_t g_{\mu\nu}\\&
+\left(p_r-p_t\right)\chi_\mu \chi_\nu +q\left(u_\mu \chi_\nu+ u_\nu \chi_\mu\right).
\end{split}
      \end{equation}
$\rho$, $p_t$,
$p_r$ and $q$ are the energy density, tangential pressure, radial pressure, and radial heat flux of the fluid respectively;  
$u^\mu=A\delta^\mu_0$ is the velocity of the fluid
and $\chi^\mu=A\delta^\mu_1$
 is a unit spacelike vector along the radial direction. 
 
 We assume the system to be self-similar of the first kind in nature i.e. the metric  admits a homothetic Killing vector \cite{maartens, maharaj1, maharaj2}. 
In the present case, we write, 
\begin{equation}
\begin{split}
 A(r,t) =r B(z),\hspace{0.2cm} \rho(r,t)=\rho(z), \hspace{0.2cm} p_r(r,t)=p_r(z),\\ p_t(r,t)=p_t(z), \hspace{0.2cm}
q(r,t)=q(z), \hspace{0.2cm}  \phi(r,t)=\phi(z),
\end{split}
\end{equation}
 where $z = \dfrac{t}{r}$.
 With this choice the field equations become self similar (i.e. the only independent variable in these equations is $z$).
 
\subsection{Field Equations}
 The Einstein field equations (in the units $8 \pi G = 1$) for the metric \eqref{metric} with energy momentum tensor \eqref{EMT} can be written as,
\begin{equation}
\label{fe1}
3\dot{A}^2-3A'^2+2AA''+\frac{4}{r}AA'={\rho}+\frac{1}{2}A^2\dot{\phi}^2+\frac{1}{2}A^2\phi'^2+V(\phi),
\end{equation}

\begin{equation}
\label{fe2}
2\ddot{A}A-3\dot{A}^2+3A'^2-\frac{4}{r}AA'=p_r+\frac{1}{2}{\phi'}^2A^2+\frac{1}{2}A^2\dot{\phi}^2-V(\phi),
\end{equation}

\begin{equation}
\begin{split}
\label{fe3}
2\ddot{A}A-3\dot{A}^2+3A'^2-\frac{2}{r}AA'-2AA''=& p_t+\frac{1}{2}A^2\dot{\phi}^2 - \frac{1}{2}{\phi'}^2A^2\\&-V(\phi),
\end{split}
\end{equation}
and
\begin{equation}
\label{fe4}
\frac{2\dot{A}'}{A}=\dot{\phi}\phi'-\frac{q}{A^2},
\end{equation}
where dot and prime denote differentiation with respect  to $t$ and $r$ respectively.

In terms of the self-similar variable $z$, equation \eqref{fe4} can be written as,
\begin{equation}
\label{fe4s}
\begin{split}
\frac{1}{B}\frac{d^2B}{dz^2}=\frac{1}{2}\left(\frac{d\phi}{dz}\right)^2+\frac{q}{2zB^2}
\end{split}.
\end{equation}
Equation \eqref{fe4s} suggests that if $\phi$ is  constant and $q=0$, $B$ will be proportional to $z$.  
Similarly, we can write down equations \eqref{fe1}-\eqref{fe3} in terms of $z$ as,
\begin{equation}
\label{fe1s}
\begin{split}
\frac{1}{B}\frac{d^2B}{dz^2}-\frac{3}{B^2}\left(\frac{dB}{dz}\right)^2+\frac{2z}{z^2-1}\frac{1}{B}\frac{dB}{dz}+\frac{1}{z^2-1}\\
=\frac{\rho}{(z^2-1)B^2}-\frac{q(1+z^2)}{2z(z^2-1)B^2}+\frac{V(\phi)}{(z^2-1)B^2},
\end{split}
\end{equation}

\begin{equation}
\label{fe2s}
\begin{split}
\frac{1}{B}\frac{d^2B}{dz^2}-\frac{3}{B^2}\left(\frac{dB}{dz}\right)^2+\frac{2z}{z^2-1}\frac{1}{B}\frac{dB}{dz}+\frac{1}{z^2-1}\\
=-\frac{p_r}{(z^2-1)B^2}+\frac{q(1+z^2)}{2z(z^2-1)B^2}+\frac{V(\phi)}{(z^2-1)B^2},
\end{split}
\end{equation}

\begin{equation}
\label{fe3s}
\begin{split}
\frac{1}{B}\frac{d^2B}{dz^2}-\frac{3}{B^2}\left(\frac{dB}{dz}\right)^2+\frac{4z}{z^2-1}\frac{1}{B}\frac{dB}{dz}-\frac{1}{z^2-1}\\
=-\frac{p_t}{(z^2-1)B^2}-\frac{q}{2zB^2}+\frac{V(\phi)}{(z^2-1)B^2},
\end{split}
\end{equation}
where   equation \eqref{fe4s} has been used to replace  terms containing derivatives of the scalar field. 
Using equations \eqref{fe1s} and \eqref{fe2s}, one can easily show that,
\begin{equation}
 \label{rhopr}
 \rho=-p_r+\frac{q(1+z^2)}{z},
\end{equation}
while \eqref{fe2s} and \eqref{fe3s} will give,
\begin{equation}\label{eqfu}
 \frac{1}{B}\frac{dB}{dz}=\frac{1}{z}+\frac{p_r-p_t}{2zB^2}-\frac{q}{2B^2}.
\end{equation}

\subsection{Conservation Equations}
We shall now write down the conservation equations for this system under the assumption that the energy momentum tensors corresponding to 
the fluid and the scalar field are conserved independently. The conservation equation for the scalar field yields the wave equation,
\begin{equation}
\label{wave}
\Box\phi-\frac{dV}{d\phi} = 0.
\end{equation}

For the present metric (\ref{metric}), this equation (\ref{wave}) translates into,
\begin{equation}
\label{wave2}
\ddot{\phi}-\phi''-2\frac{\dot{A}}{A}\dot{\phi}-2\frac{\phi'}{r}+2\frac{\phi'A'}{A}+\frac{1}{A^2}\frac{dV}{d\phi}=0.
\end{equation}

In terms of the self-similar variable $z$, we have,
\begin{equation}\label{wave3}
\frac{d^2\phi}{dz^2}-2\frac{d\phi}{dz}\Bigg[\frac{1}{B}\frac{dB}{dz}+\frac{z}{1-z^2}\Bigg]+\frac{1}{B^2(1-z^2)}\frac{dV}{d\phi} = 0.
\end{equation}          

The conservation equations for the fluid is given by,
\begin{equation}\label{consfl1}
 \nabla_\mu T_\mathrm{fluid}^{\mu\nu}=0,
\end{equation}
from which we will get two non-trivial equations,
\begin{equation}
 \label{rho}
\dot{\rho}+q^\prime=\frac{3\dot{A}}{A}(\rho+p_t)+\frac{\dot{A}}{A}(p_r-p_t)+\frac{4qA^\prime}{A}-\frac{2q}{r} ,
\end{equation}
and
\begin{equation}\label{prad}
 {p_r}^\prime+\dot{q}=\left(\frac{3A^\prime}{A}-\frac{2}{r}\right)(p_r-p_t)+\frac{A^\prime}{A}(\rho+p_t)+\frac{4q\dot{A}}{A}.
\end{equation}
For the self similar case, if we use equation \eqref{rhopr}, both equations \eqref{rho} and \eqref{prad}  yield the same equation as,
\begin{equation}\label{consfl2}
 \frac{d\rho}{dz}-z\frac{dq}{dz}=\frac{2}{B}\frac{dB}{dz}(\rho+p_t)+\frac{(1-3z^2)q}{zB}\frac{dB}{dz}+2q.
\end{equation}

\section{Raychaudhuri Equation}\label{sec3}

For a timelike congruence having velocity vector $u^\mu$, the Raychaudhuri equation is given by \cite{rc, ehlers},
\begin{equation}
\label{raych-eq}
 \frac{d\theta}{d\tau}=-\frac{1}{3}\theta^2+\nabla_{\mu}a^{\mu}-\sigma_{\mu\nu}\sigma^{\mu\nu}
 +\omega_{\mu\nu}\omega^{\mu\nu}-R_{\mu\nu}u^\mu u^\nu,
 \end{equation}
where $\theta=\nabla_\mu u^\mu$ is the expansion scalar, $\tau$ is affine parameter, 
$\sigma_{\mu\nu}=\nabla_{(\nu}u_{\mu)}-\frac{1}{3}h_{\mu\nu}\theta+a_{(\nu}u_{\mu)}$
is the shear tensor where $h_{\mu\nu}$ is the spatial metric, $\omega_{\mu\nu}=\nabla_{[\nu}u_{\mu]}-a_{[\nu}u_{\mu]}$ 
is the rotation tensor, $a^\mu=u^\nu\nabla_{\nu}u^{\mu}$ is the acceleration vector and $R_{\mu\nu}$ is the Ricci scalar.

\subsection{Focusing Condition}
We have chosen a comoving observer, so that $u^{\alpha} = A{\delta}^{\alpha}_{0}$. As the metric is conformally flat, 
the shear term and rotation term will vanish.

Now, we know that the congruence will focus \cite{poisson} within a finite affine parameter value if, 
\begin{equation}
 \frac{d\theta}{d\tau}+\frac{1}{3}\theta^2\leq 0,
\end{equation}
which, with equation \eqref{raych-eq}, leads to,
\begin{equation}\label{abcond}
 R_{\mu\nu}u^\mu u^\nu\geq \nabla_{\mu}a^{\mu}.
\end{equation}

Here the left hand side of the expression is related to the matter sector via the Einstein equations and the right hand side
is the divergence of acceleration. 
Focusing may be forbidden when the divergence of acceleration opposes the gravitational attraction and its
contribution dominates over that due to the matter part. If the divergence of acceleration term is sufficiently high, the evolution
may lead to a complete dispersal.

For the metric \eqref{metric},  the condition \eqref{abcond} in terms of the conformal factor  $A(r,t)$ is given by,
\begin{equation}\label{abcond2}
\frac{\partial}{\partial t}\left(\frac{\dot{A}}{A}\right)\geq 0.
\end{equation}

For the self similar case, the above condition \eqref{abcond2} becomes,
\begin{equation}
 \label{rccond}
\frac{d}{dz}\left(\frac{1}{B}\frac{dB}{dz}\right)\geq 0.
\end{equation}

\subsection{Raychaudhuri Equation and the dynamics of spacetime}\label{sec3.2}
 It is difficult to find exact solutions of the field equations for our system without any simplifying assumptions. 
 The Raychaudhuri equation can be applied to provide 
generic conditions, regarding the dynamics of the spacetime. These conditions may lead us to some useful 
information about the evolution. We will now try to find such conditions.

 We have the condition for focusing \eqref{rccond} from the Raychaudhuri equation as, 
\begin{equation}\label{cond1}
  \frac{d}{dz}\left(\frac{1}{B}\frac{dB}{dz}\right)\geq 0.
\end{equation}
Using \eqref{eqfu} this condition can be written as,
\begin{equation}\label{cond2}
 \frac{d}{dz}\left(\frac{p_r-p_t}{2zB^2}-\frac{q}{2B^2}\right)\geq \frac{1}{z^2}.
\end{equation}

These generic conditions like \eqref{cond1} or \eqref{cond2}, may appear useful. 
These are generic conditions if satisfied by the conformal factor or the energy momentum tensor components, will lead to the formation of a singularity. Thus, there is an important role of these conditions in dictating the dynamics of the spacetime.
Let us illustrate this point with a simple but very important example. If the fluid is a perfect isotropic fluid having no radial heat flux, the left hand side of the condition \eqref{cond2} is equal to zero. Thus, the condition is satisfied only when $z\rightarrow \infty$ which means that there will be a central singularity at $r=0$ or a singularity forms as $t\rightarrow \infty$ which is inconsequential. \\

Thus, the possibility of the formation of a singularity at a finite future or the avoidance of a central singularity will depend on pressure anisotropy or heat flux. \\

We will discuss a few more examples. The condition \eqref{cond1} or \eqref{cond2} can be recast into different forms
using the field equations. For example, if we use equation \eqref{fe4s} and \eqref{eqfu}, the condition \eqref{cond1} can be written as,
\begin{equation}\label{cond3}
 \frac{1}{2}\left(\frac{d\phi}{dz}\right)^2+\frac{q}{2zB^2}\geq \left(\frac{1}{z}+\frac{p_r-p_t}{2zB^2}-\frac{q}{2B^2}\right)^2.
\end{equation}
Now, let us consider the case where the scalar field is absent or a constant (equivalent to a cosmological constant) and $q=0$. In this case the above condition \eqref{cond3} will be satisfied only
when
\begin{equation}
 \frac{1}{z}+\frac{p_r-p_t}{2zB^2}=0,
\end{equation}
which implies $\dfrac{dB}{dz}=0$ (using equation \eqref{eqfu}) and there will  not be any evolution of the spacetime. 
Therefore, we can conclude that formation of a singularity in this case can be avoided. However,  $q\neq 0$ or $\dfrac{d\phi}{dz}\neq 0$ or both may lead to the formation of a singularity. If we have a large rate of change of the scalar field, compared to the other terms, present in the condition \eqref{cond3}, singularity formation is inevitable. \\

Let us discuss another example where $\dfrac{d\phi}{dz}=0$ and $p_r=p_t$. The condition \eqref{cond3} then translates into,
\begin{equation}
 \frac{q^2}{4 B^4}-\frac{3q}{2zB^2}+\frac{1}{z^2}\leq 0,
\end{equation}
which gives,
\begin{equation}\label{constraint}
 \frac{(3-\sqrt{5})B^2}{z}\leq q \leq \frac{(3+\sqrt{5})B^2}{z}.
\end{equation}
Thus, there will be singularity formation only when the heat flux of the fluid satisfies this constraint \eqref{constraint}.

\subsection{Exact solutions and the Raychaudhuri equation}\label{sec3.3}
We have not used exact solutions of the field equations in our discussion so far. It is worthwhile to check the consistency of
the conclusions arrived at using the Raychaudhuri equation with those using exact solutions whenever available. 
For many special cases, i.e., with various sources, exact solutions for the metric can be found out. Some of them are listed in the  \ref{appen}. With an exact solution, one can explicitly find  whether  there is a collapse or an expansion and whether the collapse, if there is any, results in a singularity. We have checked that in all such cases the conclusions are consistent with those obtained from the focusing condition (\ref{rccond}) found out from the Raychaudhuri equation. One such nontrivial example is described in detail below. \\

We will discuss the case where a scalar field and a fluid both are present and we make an assumption that $p_r\neq p_t$ but $q=0$. 
For details, see \ref{fifth}. The solution for $B$ and $A$ in this case are respectively given by,
\begin{equation}\label{Bsolprop}
 B=\left(Fz^{2\beta}-k\right)^\frac{1}{2\beta},
\end{equation}
and
\begin{equation}
 A=\left(Ft^{2\beta}-kr^{2\beta}\right)^\frac{1}{2\beta},
\end{equation}
where $\beta$ is a constant given by $p_t = \beta p_r.$
For $\beta>0$, there may be a zero proper volume singularity when $t\rightarrow\infty$ and/or $r\rightarrow\infty$, 
which can be excluded from the discussion.
There will be a dispersal in this case
at $z=\left(\dfrac{k}{F}\right)^{\frac{1}{2\beta}}$. 
When
$\beta<0$, formation of a singularity 
at  $z=\left(\dfrac{k}{F}\right)^{\frac{1}{2\beta}}$ is inevitable. In this case, a dispersal occurs at $t=0$. There may be a dispersal at $r=0$ 
which is not physically significant.\\

Using the solution for $B$ (equation \eqref{Bsolprop}),
one can show,
\begin{equation}\label{rcprop}
 \frac{d}{dz}\left(\frac{1}{B}\frac{dB}{dz}\right)=-\frac{2F^2\beta z^{(4\beta-2)}}{\left(F z^{2\beta}-k\right)^2}
 +\frac{F(2\beta-1) z^{(2\beta-2)}}{\left(F z^{2\beta}-k\right)}.
\end{equation}
When $z\rightarrow\left(\dfrac{k}{F}\right)^{\frac{1}{2\beta}}$, the first term in the 
right hand side of the above expression dominates. Only when $\beta<0$, the condition \eqref{cond1}, obtained using the Raychaudhuri equation, is fulfilled in this region and a singularity 
within finite $z$ is unavoidable. On the other hand if $\beta>0$, a dispersal will occur in this region. Singularity formation  as $z\rightarrow\infty$ is a possibility in this case.
For $\beta<0$, it is easy to show that the right hand side of the equation \eqref{rcprop} must be negative as $z\rightarrow 0$
which makes a dispersal possible here. \\

\subsection{Critical Phenomena and the Raychaudhuri equation}\label{sec3.4}
From the exact solutions in  \ref{appen}, we have found that collapse or dispersal correspond to the  the quantity
$\dfrac{d}{dz}\left(\dfrac{1}{B}\dfrac{dB}{dz}\right)$ 
being positive or negative, respectively. Therefore, a transition from collapse to dispersal or vice versa will be accompanied with
a change of the sign of this quantity. Thus, a relation between  focusing condition and the critical phenomena seems to be indicated. We will illustrate this point using the example discussed
in the previous section.

In this example, let us take the $z\rightarrow\left(\dfrac{k}{F}\right)^{\frac{1}{2\beta}}$ into consideration. When $\beta$ is negative a singularity forms at $z=\left(\dfrac{k}{F}\right)^{\frac{1}{2\beta}}$ while dispersal occurs when $\beta$ is positive. Thus, $\beta$ can be treated as a critical parameter. We have seen in the previous section that positive or negative $\beta$ corresponds to the situation when the quantity $\dfrac{d}{dz}\left(\dfrac{1}{B}\dfrac{dB}{dz}\right)$ is negative or positive respectively as $z\rightarrow\left(\dfrac{k}{F}\right)^{\frac{1}{2\beta}}$. This shows that the critical parameter $\beta$ determines the signature of $\dfrac{d}{dz}\left(\dfrac{1}{B}\dfrac{dB}{dz}\right)$ which in turn determines the focusing (collapse) or dispersal of the spherically symmetric distribution and the critical value of $\beta$ is zero.

\section{Conclusion}\label{sec4}

We have found the focusing condition \eqref{cond1} for a  self similar  matter distribution in a conformally flat spherically symmetric spacetime. Although the spacetime, we have considered, has a stringent symmetry requirement, the matter distribution is quite general and includes anisotropic fluid pressure, heat flux, as well as a minimally coupled scalar field. The nature of the conclusions drawn from this condition is verified against quite a few exact solutions that are available, and one of them is discussed in detail. \\

Thanks to the investigations as in references \cite{chop, bcg, gundprl} and others, it is now well known that a scalar field collapse may have a critical phenomenon associated with it. The condition developed from the Raychaudhuri equation helps visualizing the critical phenomena in general. This remains valid even when the matter distribution includes a fluid. We have discussed one example, where exact solution is available, in detail. In this example, we could see the existence of a critical parameter, and could also determine the  critical value ($\beta = 0$). It is intriguing to note that the parameter has nothing to do with the scalar field in this case and is determined only by the fluid, as $\beta$ is simply a parameter connecting the radial and transverse fluid pressure. With the help of this example we have explicitly discussed the relation between the focusing condition and the critical phenomena.

\begin{acknowledgements}
Shibendu Gupta Choudhury (SGC) thanks Council of Scientific and Industrial Research, India for the financial support.
SGC would like to thank Prof. Sayan Kar, Indian Institute of Technology, Kharagpur for guidance and useful
discussions.
\end{acknowledgements}

\appendix

\section{A Few Exact Solutions}\label{appen}

Some examples of exact solutions with the assumption of self similarity, for a conformally flat, spherically symmetric spacetime in presence of a scalar field and a fluid, imperfect in general, are given here. Some of the solutions included are already there in the literature.

\subsection{Massless scalar field}\label{massless}
 Let us first consider the simplest case where the contribution to the matter part comes only from a massless scalar field. From the equation \eqref{eqfu} we have, 
\begin{equation}\label{mlu}
 \frac{1}{B}\frac{dB}{dz}=\frac{1}{z},
\end{equation}
which gives,
\begin{equation}\label{solB}
 B=m z,
\end{equation}
where $m$ is  constant of integration. Thus, the solution for $A(r,t)$ is, 
\begin{equation}\label{solA}
 A=rB=m t.
\end{equation}
But we should note that with equation \eqref{mlu}, equations \eqref{fe1s}, \eqref{fe2s} and \eqref{fe3s} all yield,
\begin{equation}
\frac{3}{z^2(z^2-1)}=0,
\end{equation}
 which is inconsistent for finite $z$ values. Thus, a consistent solution of the field equations with a massless scalar field
as matter source is not possible 
under the mentioned assumptions.

\subsection{Scalar field with a non-zero potential}\label{massive}
Even if we include a potential in the energy momentum tensor of the scalar field,
we have the same solutions for $B$ and $A$ as in the previous case (equations \eqref{solB} and \eqref{solA} respectively).
Here with equation \eqref{solB}, equations \eqref{fe1s}, \eqref{fe2s} and \eqref{fe3s} all yield,
\begin{equation}
 V(\phi)=3m^2.
\end{equation}
Therefore, the potential must remain constant.

From equation \eqref{fe4s} we have,
\begin{equation}
 \phi=\text{ constant},
\end{equation}
which is consistent with the equation \eqref{wave3}. Thus, the net energy momentum tensor effectively behaves like that of a cosmological constant. 

From equation \eqref{solA}, we can conclude that there is a singularity only when $t$ approaches infinity where the scale factor 
(inverse of $A$) becomes zero which signifies an ever
collapsing solution.
At $t=0$, $A=0$ i.e. the scale factor becomes infinite which signifies a dispersal.

\subsection{Scalar field along with a perfect fluid}\label{perfect}
If we include perfect isotropic fluid ($p_r=p_t$ and $q=0$) along with the scalar field, 
the solution for $B$ will not change as dictated by equation \eqref{eqfu}.
From equation \eqref{rhopr} we have,
\begin{equation}
 \rho=-p.
\end{equation}
$\rho$, $p$, $\phi$ and $V(\phi)$ will remain constant in this case. Thus the matter turns out to be
the cosmological constant for a consistent solution. Clearly, the conclusions in this case remain the same as in the previous case.

 In the examples, discussed so far we have not observed any possibilities of singularity formation within finite time and the matter turns out to be the cosmological constant. 

\subsection{Fluid with isotropic pressure and radial heat flux}\label{fourth}

This particular case, with $p_r=p_t$ and $q\neq 0$, has already been studied by Chan, Silva and Rocha \cite{chan0}. They found that there is a singularity formation due to collapse as $t\rightarrow 0$. In this case from equations \eqref{eqfu} and \eqref{fe4s} we have,
\begin{equation}
 \frac{z}{B}\frac{d^2B}{dz^2}+\frac{1}{B}\frac{dB}{dz}-\frac{1}{z}=0.
\end{equation}
The solution for $B$ is given by,
\begin{equation}\label{s1B}
 B=\frac{C z^2+2D}{2z},
\end{equation}
where $C$ and $D$ are  constants of integration.
 From the above equation \eqref{s1B} we have,
\begin{equation}
 A=\frac{Ct^2+2Dr^2}{2t}.
\end{equation}
The expression for $A$ confirms that there is a zero proper volume singularity at $t=0$. The scale factor also becomes zero when $t\rightarrow \infty$ and/or $r\rightarrow \infty$. These possibilities are not worth considering.
There may be a dispersal
at $z^2=\dfrac{t^2}{r^2}=-\dfrac{2D}{C}$ only if $C$ and $D$ are of opposite signs.

\subsection{Scalar field along with an anisotropic fluid}\label{fifth}
In this case, we have $p_r\neq p_t$ but $q= 0$. Now, from \eqref{rhopr} and \eqref{eqfu} we have,
\begin{equation}\label{cc}
 \rho=-p_r,
\end{equation}
and
\begin{equation}\label{Bprop}
 \frac{1}{B}\frac{dB}{dz}=\frac{1}{z}+\frac{p_r-p_t}{2zB^2}.
\end{equation}
The same equation of state as in \eqref{cc} was obtained by Brandt {\it et al} \cite{chan2} from the requirement of self-similarity even with a relaxation of the conformally flat condition.
In this case, the conservation equation for the fluid (equation \eqref{consfl2}) takes the form,
\begin{equation}\label{consprop}
 \frac{dp_r}{dz}=\frac{2}{B}\frac{dB}{dz}(p_r-p_t).
\end{equation}

It is difficult to find a solution for the conformal factor in this case without further simplifications. Thus, we will assume that the tangential pressure
is proportional to radial pressure i.e. $p_t=\beta p_r$. Brandt {\it et al} worked with a similar assumption for a spherically symmetric spacetime which is not conformally flat \cite{chan1}. With this assumption, solving equation \eqref{consprop} we have,
\begin{equation}
 p_r=n B^{2(1-\beta)},
\end{equation}
where $n$ is  constant of integration.
If we replace this in equation \eqref{Bprop}, the solution for $B$ comes out as,
\begin{equation}\label{Bsolpropap}
 B=\left(Fz^{2\beta}-k\right)^\frac{1}{2\beta},
\end{equation}
where $F$ is constant of integration and $k=\dfrac{n(1-\beta)}{2}$.
 Consequently, the solution for $A$ is,
\begin{equation}
 A=\left(Ft^{2\beta}-kr^{2\beta}\right)^\frac{1}{2\beta}.
\end{equation}\\
From equation \eqref{cc} one can see that
the radial pressure of the fluid comes out to be negative. For consistency, we may consider the fluid to be a viscous fluid.\\


\begin{thebibliography}{99}
\bibitem{datt} B. Datt, Z. Phys. {\bf 108}, 314 (1938)
\bibitem{os} J. R. Oppenheimer, H. Snyder, Phys. Rev. {\bf 56}, 455 (1939)
\bibitem{joshi1} P. S. Joshi, {\it Global aspects in Gravitation and Cosmology}, International Series of Monographs on Physics (Clarendon Press (OUP), Oxford, 1993)
\bibitem{joshi2} P. S. Joshi, Pramana {\bf 55}, 529 (2000)
\bibitem{christo} D. Christodoulou, Commun. Math. Phys. {\bf 109}, 591 (1987)
\bibitem{christo2} D. Christodoulou, Commun. Math. Phys., {\bf 109}, 613 (1987)
\bibitem{christo3} D. Christodoulou, Ann. Math. {\bf 140}, 607 (1994)
\bibitem{chop} M. W. Choptuik, Phys. Rev. Lett. {\bf 70}, 9 (1993)
\bibitem{bcg} P. R. Brady, C. M. Chambers, S. M. C. V. Goncalves, Phys. Rev. D {\bf 56}, R6057
(1997)
\bibitem{gundprl} C. Gundlach, Phys. Rev. Lett. {\bf 75}, 3214 (1995)
\bibitem{ola} I. Olabarrieta, J. F. Ventrella, M. W. Choptuik, W. G. Unruh, Phys. Rev. D {\bf 76},
124014 (2007)
\bibitem{olachop} I. Olabarrieta, M. W. Choptuik, Phys. Rev. D {\bf 65}, 024007 (2001)
\bibitem{brady} P. R. Brady, Class. Quant. Grav. {\bf 11}, 1255 (1995)
\bibitem{Gundlach1} C. Gundlach, J.M. Martin-Garcia, Liv. Rev. Rel. {\bf 10}, 5 (2007)
\bibitem{rc} A. K. Raychaudhuri, Phys. Rev. {\bf 98}, 1123 (1955), reprinted as a `Golden Oldie', Gen. Relativ. Gravit. {\bf 32}, 749 (2000)
\bibitem{ehlers} J. Ehlers, Akad. Wiss. Lit. Mainz, Abhandl. Math.-Nat. Kl. {\bf 11}, 793 (1961), translation: J. Ehlers, Gen. Relativ. Gravit. {\bf 25}, 1225 (1993)
\bibitem{kt} A. P. Kouretsis, C. G. Tsagas, Phys. Rev. D {\bf 82}, 124053 (2010)
\bibitem{som} M. M. Som, N. O. Santos, Phys. Lett. A {\bf 87}, 89 (1981)
\bibitem{maiti} S. R. Maiti, Phys. Rev. D {\bf 25}, 2518 (1982)
\bibitem{modak} B. Modak, J. Astrophys. Astron. {\bf 5}, 317 (1984)
\bibitem{bhui} A. Banerjee, S. B. D. Choudhury, B. K. Bhui, Phy. Rev. D {\bf 40}, 670 (1989)
\bibitem{patel} L. K. Patel, R. Tikekar, Mathematics Today {\bf IX}, 19 (1991)
\bibitem{schafer} D. Schafer, H. F. Goenner, Gen. Relativ. Gravit. {\bf 42}, 2119 (2000)
\bibitem{ivanov} B. V. Ivanov, Gen. Relativ. Gravit. {\bf 44}, 1835 (2012)
\bibitem{herrera} L. Herrera, G. Le Denmat, N. O. Santos, A. Wang, Int. J. Mod. Phys. D {\bf 13}, 583 (2004)
\bibitem{hamid} A. I. M. Hamid, R. Goswami, S. D. Maharaj, Class. Quant. Grav. {\bf 31}, 135010 (2014)
\bibitem{scnbepjc} S. Chakrabarti, N. Banerjee, Eur. Phys. J. C {\bf 77}, 166 (2017)
\bibitem{scnbprd} N. Banerjee, S. Chakrabarti, Phys. Rev. D {\bf 95}, 024015 (2017)
\bibitem{chan0} R. Chan, M. F. A. da Silva, J. F. V. da Rocha, Int. J. Mod. Phys. D {\bf 12}, 347 (2003)
\bibitem{brandt} C. F. C. Brandt, L. M. Lin, J. F. V. da Rocha, A. Z. Wang, Int. J. Mod. Phys. D {\bf 11}, 155
(2002)
\bibitem{chan1} C. F. C. Brandt, M. F. A da Silva, J. F. V. da Rocha, R. Chan, Int. J. Mod. Phys. D {\bf 12},
1315 (2003)
\bibitem{chan2} C. F. C. Brandt, R. Chan, M. F. A da Silva, J. F. V. da Rocha, Int. J. Mod. Phys. D {\bf 15},
1407 (2006)
\bibitem{chan3} R. Chan, M. F. A. da Silva, C. F. C. Brandt, Int. J. Mod. Phys. D {\bf 23},
1450056 (2014)
\bibitem{maartens} R. Maartens, S.D. Maharaj, Class. Quant. Grav. {\bf 3}, 1005 (1986)
\bibitem{maharaj1} S. Moopanar, S. D. Maharaj, J. Eng. Math. {\bf 82}, 125 (2013)
\bibitem{maharaj2} S. Moopanar, S. D. Maharaj, Int. J. Theor. Phys. {\bf 49}, 1878 (2010)
\bibitem{poisson} E. Poisson, {\it A Relativist's Toolkit: The Mathematics of Black-Hole Mechanics} (Cambridge University Press, Cambridge 2004)

\end{thebibliography}
\end{document}